\documentclass[5p]{elsarticle} 

\usepackage{amsmath}
\usepackage{amssymb} 
\usepackage{graphicx}      
\usepackage{natbib}        
\usepackage[utf8]{inputenc}
\usepackage{graphicx}
\usepackage{color,soul}
\graphicspath{ {} }
\usepackage{float} 
\usepackage{textcomp, gensymb}
\usepackage{placeins}

\journal{arXiv}

\begin{document}

\begin{frontmatter}

\title{ROM-Based Stochastic Optimization for a Continuous Manufacturing Process}

\author[label1]{Raul Cruz-Oliver}

\author[label2]{Luis Monzon}

\author[label3]{Edgar Ramirez-Laboreo\corref{cor1}}
\ead{ramirlab@unizar.es}

\author[label2]{Jose-Manuel Rodriguez-Fortun}

\affiliation[label1]{organization={ETH Zurich},
            city={Zurich},
            postcode={8092},
            country={Switzerland}}

\affiliation[label2]{organization={ROMEM Research Group, Instituto Tecnologico de Aragon (ITA)},
            city={Zaragoza},
            postcode={50018},
            country={Spain}}

\affiliation[label3]{organization={Departamento de Informatica e Ingenieria de Sistemas (DIIS) and Instituto de Investigacion en Ingenieria de Aragon~(I3A),\\Universidad de Zaragoza},
            city={Zaragoza},
            postcode={50018},
            country={Spain}}

\cortext[cor1]{Corresponding author}

\begin{abstract}
This paper proposes a model-based optimization method for the production of automotive seals in an extrusion process. The high production throughput, coupled with quality constraints and the inherent uncertainty of the process, encourages the search for operating conditions that minimize nonconformities. The main uncertainties arise from the process variability and from the raw material itself. The proposed method, {which is} based on Bayesian optimization, takes these factors into account and obtains a robust set of process parameters. Due to the high computational cost and complexity of performing detailed simulations, a reduced order model is used to address the optimization. The proposal has been evaluated in a virtual environment, {where it has been verified that it is able to minimize the impact of process uncertainties. In particular, it would significantly improve the quality of the product without incurring additional costs, achieving a 50\% tighter dimensional tolerance compared to a solution obtained by a deterministic optimization algorithm.}
\end{abstract}


\end{frontmatter}

\section{Introduction}

Improving manufacturing processes in industry requires balancing production speed and quality standards in a changing environment. This uncertainty and the increasing complexity of production systems represent fundamental challenges for the new paradigms of Industry 4.0 and 5.0~\cite{Bakon2022} and practical methods such as model-based optimization are crucial to address them. The objective is to identify reliable process parameters that robustly meet quality standards.

The optimization of uncertain industrial processes is a major research activity due to its impact on quality, efficiency, and cost. The authors of~\cite{Bakon2022} consider three sources of uncertainty: the variability of internal and external factors influencing the process, such as uncontrolled events, phenomena, or human behavior; the knowledge uncertainty associated with parameter tolerances or model errors; and the decision uncertainty caused by changes in the operation policies inside or outside the process. To address these issues, the authors propose proactive and reactive strategies to improve initial predictive solutions based on static optimizations. The proactive approach considers the uncertainty in the optimization, while the reactive approach modifies an initial optimization in response to events. Similarly, the authors of~\cite{Chen2018} distinguish three proactive approaches for the optimization of chemical processes: first, robust optimization, which takes into consideration the worst-case scenario; second, stochastic optimization, which optimizes the expectation of the outcome; and finally, chance-constrained programming, which allows some constraint violations in a robust optimization.

The choice of an optimization strategy is related to the knowledge of the system and the mathematical model used to represent it~\cite{Misener2023}. The complexity of new production systems has fostered the development of specific solutions for both identification and modeling. The model can be purely data driven, physically based, or a hybrid combining the previous two~\cite{Kasilingam2024}. Apart from that, a single representation can be used for the whole process or it can be divided into different levels with separate surrogate representations. In all cases, in order to be robustly used for optimization, the model must be accurate and avoid overfitting problems, which would make it sensitive to uncertain data and prone to prediction errors when extrapolating~\cite{Misener2023}.

There are many examples of pure data-driven representations in the literature. The work in~\cite{Herceg2023} compares the use of dynamic long short-term memory (LSTM) with more traditional solutions for an industrial isomerization process. The review work in~\cite{MUNIR20210725} describes various linear and nonlinear regression methods, such as principal component analysis (PCA), artificial neural networks (ANN), or extreme learning machines (ELM), for monitoring and controlling a hot-melt extrusion process in the pharmaceutical industry. Also in this industrial field are the models summarized in~\cite{Dong2023}. They range from pure statistical identification, such as design of experiments (DoE) or Bayesian inference, to machine learning (ML) models, such as support vector machine (SVM), random forest (RF), and deep learning models. A different approach is described in~\cite{Lambard2022}, where the authors use RF models in the framework of ALMLBO---Active Learning Assisted by Machine Learning and Bayesian Optimization---for an extrusion process of \mbox{Nd-Fe-B} magnets. The authors in~\cite{Echeverria2024}, on the other hand, propose a scalable method for modeling continuous systems by mixing Gaussian models identified using Dirichlet process clustering.

Some examples can also be found using purely physically based representations. For instance, {the work in~\cite{Cegla2021}} describes a detailed temperature and pressure model for the reactive extrusion of $\epsilon-$Caprolactone. Discrete event descriptions are also typically used in manufacturing processes, as in~\cite{Zhao2022}, where a mobile manufacturing workshop is modeled using simplified parameters per stage. For polymer extrusion processes, the review in~\cite{Nastaj2021} summarizes different approaches using detailed finite element (FE) and computational fluid dynamics (CFD) models combined with optimization techniques, such as genetic algorithms, for process optimization and scale-up.

Finally, the combination of data-driven with physical descriptions in hybrid models {overcomes the limitations of the former, which usually suffer from small amounts of data, especially under faulty conditions, and the simplifications usually necessary in physically based modeling~\cite{Kasilingam2024}. This hybrid approach can be done in different ways}, as described in~\cite{Sharma2022}, where the authors use several examples from the chemical industry to illustrate what they call science-guided machine learning (SGML). They consider two approaches: ML models complementing physically-based models, and vice versa. In the first case, ML is used to enrich some parts of a physically based model, to obtain reduced order models (ROM) from virtual data, or even for identifying a physical law from data. In the second case, the structure of the ML model is defined considering the underlying physical phenomena. The first approach is used in~\cite{HAMID2022100044}, where a ROM based on ANNs is obtained from a highly detailed physical model obtained with the software iCON-Symmetry. A similar approach for a plastic extrusion process is described in~\cite{SARISHVILI_20211110}, where experimental and virtual data from a detailed physico-chemical model are combined to train a stacked autoencoder (SAE) network. A further work in~\cite{BURR_20230405} uses inversion of the SAE classifier to estimate the parameters required in the extrusion process. 

{The present paper describes a parameter optimization method for a continuous polymer extrusion line that produces water and sound insulation door seals for the automotive industry.} The high throughput of the line combined with the inherent uncertainty in the process {requires optimized parameters which can successfully deal with the inherent variability}. They should properly work under a wide range of conditions to avoid later reconfiguration as changing parameter definitions with continuous adjustments would result in economic and quality losses. It is important to note that most of the processes involved are of thermal nature and thus require significant time to reach a steady state. This implies that reconfigurations in the line take considerable time be effective.  

The methodology presented in this article proposes the use of a reduced order hybrid model obtained with TWINKLE~\cite{ZAMBRANO2020100419} combined with Bayesian optimization to find an optimal set of parameters in the presence of the system uncertainties. The ROM condenses the information from detailed physics-based representations implemented in FE software over the entire operating range. In addition to that, it has a simple structure with a minimum number of parameters, which allows its efficient use for stochastic optimization and reduces the risk of overfitting. On the other hand, Bayesian optimization has been chosen as the optimization strategy due to its implicit consideration of the uncertainties in the system under evaluation~\cite{Shahriari2016taking,frazier2018bayesian}. To highlight the effect of its variable nature, the optimization has also been conducted with a deterministic simplex approach~\cite{lagarias1998convergence} for comparison. Nevertheless, the goal is not to compare several optimization algorithms, but only to expose the nature of the optimization function. Additionally, we would like to remark that, besides the presented method or modifications of it, other approaches could also be used depending on the final applications, such as genetic algorithms or reinforcement learning, among others. The review work presented in~\cite{Elaziz2020}, for instance, summarizes several metaheuristic models that have been successfully used in the design and optimization of mechanical systems.

The paper is organized as follows. First the extrusion system is described in Section~\ref{sec:sys_desc}. After that, both the reference FE model and the ROM are presented in Section~\ref{sec:model}, also including the uncertainty representations considered. Section~\ref{sec:optim} describes the optimization problem and the results are summarized in Section~\ref{sec:results}. Final conclusions appear in Section~\ref{sec:conclusions}.

\section{System description and optimization strategy}
\label{sec:sys_desc}

The current section describes the production line, as well as the methodology followed to perform the optimization in a reasonable time scale.

\subsection{Process description}

\begin{figure}[t]
    \centering
	\includegraphics[width=85mm]{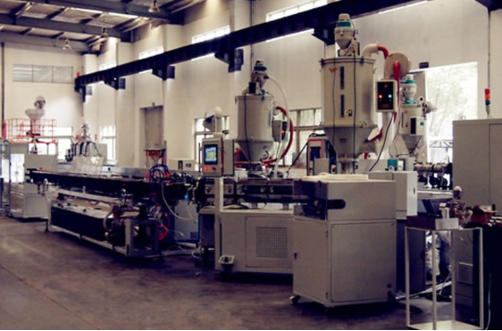}
	\caption{{Extrusion line}}
	\label{extrussion_line}
\vspace{\floatsep}
    \centering
	\includegraphics[width=65mm]{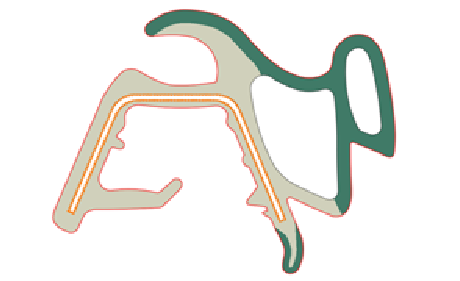}
	\caption{Automotive seal produced in the extrusion line}
	\label{part_image}
\vspace{\floatsep}
    \centering
	\includegraphics[width=\linewidth]{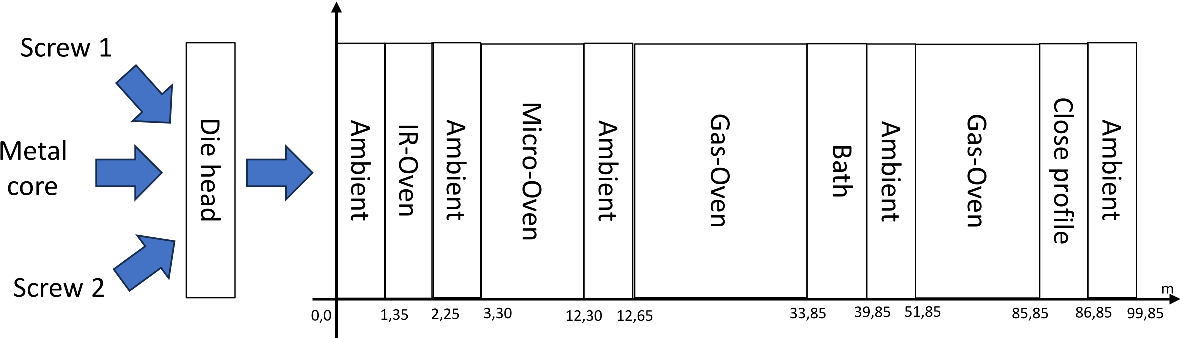}
	\caption{{Detail of the stages in the extrusion process}}
	\label{extrussion_stages}
\end{figure}

As stated, the present work is focused on a continuous extrusion process for automotive seals {(Figure~\ref{extrussion_line})}. These parts, depicted in Figure \ref{part_image}, {are made of an ethylene-propylene-diene monomer (EPDM) rubber and a metal core}.  The process is divided into two main phases. In the first one, two extrusion screws feed the raw material through a die together with the metal core. The second phase involves the sequence of thermal treatments listed in Figure~{\ref{extrussion_stages}}. During this phase, the curing and foaming processes of the material take place.  {The product is moving along a series of stages which alternates heating and cooling processes, with long infrared, microwave and gas-ovens separated by short ambient cooling stages. The gas ovens are more than 20~m long and have in consequence long thermal inertia.} These thermal processes are the subject of the optimization described in the next section. The heating and cooling treatments of this second phase can be summarized as follows:
\begin{itemize}
	\item Ambient cooling after extrusion.
	\item Infrared oven.
	\item Ambient cooling.
	\item Microwave oven.
	\item Ambient cooling.
	\item Convective gas oven.
	\item Cooling bath.
	\item Ambient cooling.
	\item Convective gas oven.
    \item The profile is closed.
	\item Final ambient cool-down.
\end{itemize}

The production line is equipped with low-level control systems implemented on industrial PCs together with a SCADA (Supervisory Control And Data Acquisition) system {{\cite{boyer2009scada}}, which allows for a remote high-level monitoring of the entire control architecture using a graphical user interface.} Quality control is performed on samples collected in the laboratory. 

\subsection{Model-based process optimization} 

The system under study is a continuous operating line with stringent requirements both in terms of throughput and quality. Given the system characteristics, it is completely unfeasible to perform the parameter optimization acting directly on the real system, using quality control data as the only source of information. Apart from the dangerous conditions and potential infrastructure damage, the operation would be extremely long due to the time required for the line to reach stationary conditions and the sampled-based quality control to feed the result back. Thus, it is clearly advantageous to have a model to predict the output of the system in stationary conditions and use it in an offline-fashion optimization. This is not an easy task due to the many factors affecting the quality result, such as the curing and foaming processes and the inherent uncertainty in the process parameters (temperature distribution in the ovens and material properties, among others). 

\begin{figure}[t]
    \centering
	\includegraphics[width=\linewidth]{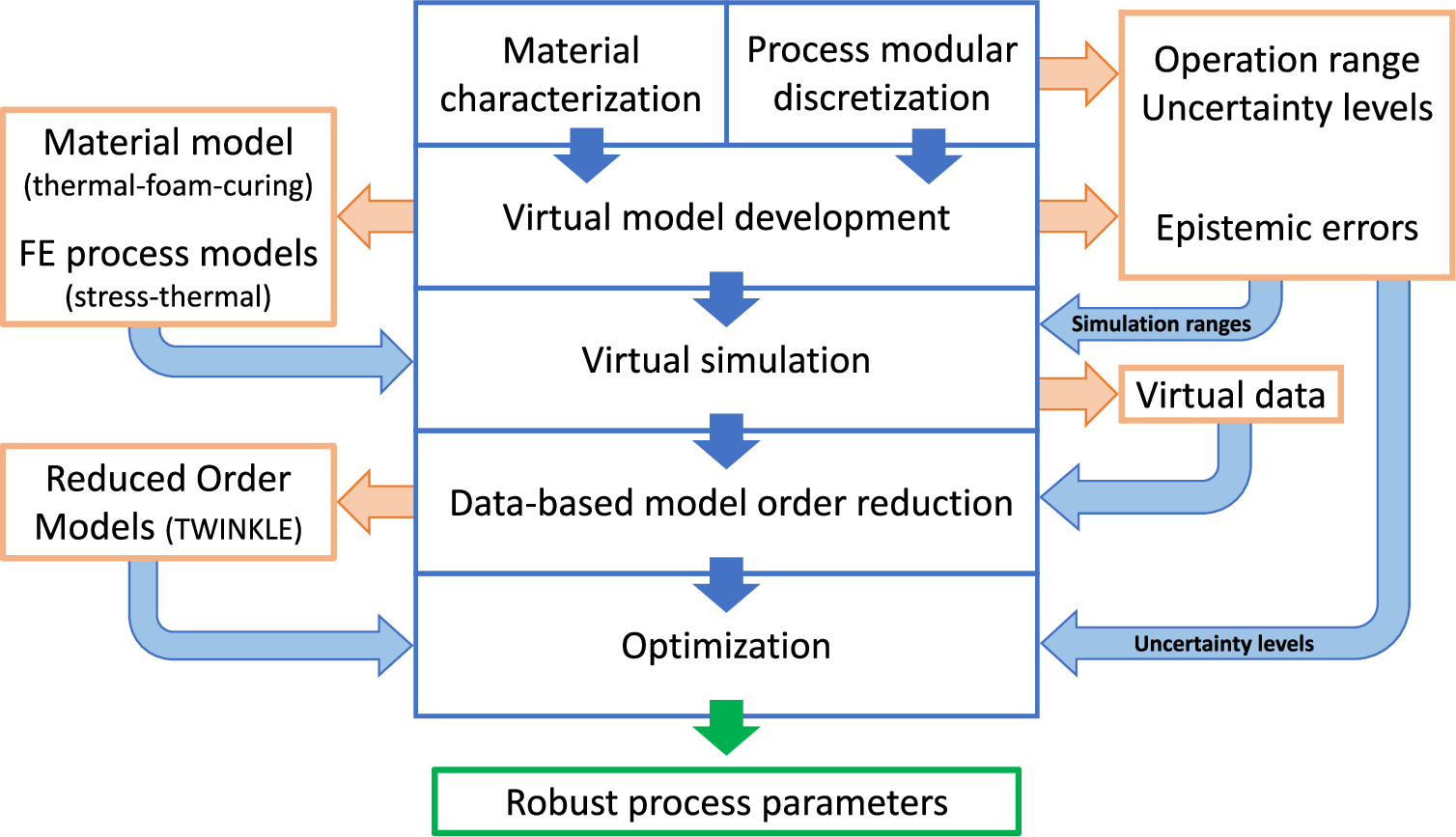}
	\caption{{Methodology outline. The blue and orange boxes correspond, respectively, to steps in the process and intermediate results. The result (green box) is the set of robust process parameters.}}
	\label{method_image}
\end{figure}

The approach proposed in this paper combines machine learning and detailed physics-based descriptions. A ROM is used to reduce the complexity of a comprehensive FE description of the extrusion line, including detailed material characterization results that take into account the vulcanization and foaming processes. The detailed FE description of the process, which is implemented in ABAQUS, was described in previous works~\cite{Viejo2022,polym14061101}. Unfortunately, this model cannot be solved time-efficiently during optimization, which is why it is necessary to use the reduced-order model built using virtual data generated by the FE model. This ROM was described in~\cite{Monge2019ReducedOM} as well as in~\cite{Viejo2022}, where a sensitivity analysis of the different parameters of the process on the final quality of the product was also performed. The surrogate model from the detailed FE model was obtained using the tensor factorization tool TWINKLE~\cite{ZAMBRANO2020100419}. In the present work, that ROM obtained with TWINKLE is used to optimize the process parameters in order to improve the final quality of the product. For this purpose, a Bayesian optimization method is used, as it takes into account the uncertainty of the process and the epistemic errors of the models used to perform the optimization~\cite{Shahriari2016taking,frazier2018bayesian}. Figure~\ref{method_image} summarizes the methodology used in this work.

For constructing the ROM, the main parameters describing the material and the process were condensed considering variability levels for process and material uncertainty, and also taking into account epistemic errors {arising from modeling simplifications}, such as the microwave heating phenomena. {The variability ranges considered in the system, which were also determined in the previously cited works based on experimental tests and benchmark simulations with the models, are summarized in Table~\ref{ref_values}}. The table distinguishes between uncertain (unc.) and deterministic (det.) parameters. The former are those that cannot be accurately controlled or that correspond to physical processes whose description in the model is uncertain. {The vector $u \in \mathbb{R}^3$ will be used to denote the array formed by these three parameters. On the other hand, the deterministic ones correspond to the process parameters that can be accurately controlled. Mathematically, these will be represented by means of the vector $c \in \mathbb{R}^6$.}  The parameters include those from the different process phases: extrusion speed through the die, pressure in the two cavities (corresponding to the two extruded materials), ratio of RPM related with the different screw speeds for both materials in order to compensate the different foaming behavior, heat in infrared and microwave ovens (ratio with respect to the nominal value), and temperature in the two gas ovens. The stochastic nature of the extrusion speed and the nominal heat in the microwave results from the process uncertainty and epistemic errors from its simulation in the detailed FE model.

\begin{table}[t]
    \centering
    \begin{tabular}{ |l|c|c| } 
        \hline
        Parameter & Type & Value $\pm$ range\\
        \hline
        Extrusion speed & unc. & $20 \pm 5$ m/min\\
        Foaming expansion coeff. & unc. & $0.1275 \pm 0.1025$\\
        Nom. heat in microw. oven & unc. & $0.55 \pm 0.45$\\
        Pressure in big cavity & det. & $1500 \pm 300$ Pa\\
        Pressure in small cavity & det. & $400 \pm 300$ Pa\\
        Ratio of RPM & det. & $0.335 \pm 10\%$\\
        Nom. heat in infrared oven & det. & $0.95 \pm 0.15$\\
        Temperature in gas oven (1) & det. & $380 \pm 100 \degree $C\\
        Temperature in gas oven (2) & det. & $350 \pm 100 \degree $C\\
        \hline
    \end{tabular}
    \caption{Process parameters (unc.: uncertain; det.: deterministic)}
    \label{ref_values}
\end{table}

\section{Modeling}  
\label{sec:model}

\subsection{Finite element model} 

The simulation of the continuous extrusion manufacturing process has been performed with a multi-physics FE model~\cite{Viejo2022}, which describes the rubber transformation through the different ovens. The simulation considers 11 steps corresponding to those in Figure \ref{extrussion_line}. The heating and cooling sources change in the simulation according to the process phase. The boundary conditions in the model simulate the extruded profile either suspended or resting on supporting elements. The metal core is co-extruded at a certain speed but the model contemplates the relative movement between the material and the core. The infrared oven is described as a superficial flux. After that, the microwave oven is simulated as a volumetric flux. For the gas ovens and the ambient cooling phases, the heat exchange is simulated as a convective process. 

One important aspect to take into account during the simulation is the coupling between the kinetic, thermal and mechanical fields. The thermal-stress relation is greatly affected by the foaming and vulcanization processes. This is achieved by means of different subroutines in the ABAQUS environment modeling the behaviour of the material, which has previously been experimentally characterized. The details of the material model appear in~\cite{polym14061101}. It is described in the mechanical domain as linear elastic given the low strain levels during the process. The representation also includes the expansion due to thermal loads and foaming process. The variability of the material properties due to foaming is described by using the Mori-Tananka approach~\cite{2013_Ghezal}. The thermal field includes the dependency on temperature, as well as foaming and curing degrees, whose dynamics are described with the Kamal-Sourour reaction model~\cite{1976_Kamal}.

The reference model is 2.5D due to a pre-strain in the longitudinal direction which is introduced to compensate the expansion caused by the foaming process.

\subsection{Reduced order model (ROM)}  

The previous detailed model is computationally complex. Therefore, in order to run the optimization in an efficient way, it is necessary to have a simpler representation. To do that, a ROM {specifically designed for predicting the deformation of the section} was fitted using the TWINKLE library~\cite{ZAMBRANO2020100419}. The reference dataset consists of virtual results in stationary conditions from simulations with the detailed model in the parameter ranges shown in Table~\ref{ref_values}.

The objective of the ROM is to model the deformation of the cross section of the material along the line, both in the x-axis and in the y-axis, in a computationally efficient way. That is, to estimate
\begin{equation}
    {
    \left(\begin{array}{c}
       \Delta x^{[i]}(z) \\
       \Delta y^{[i]}(z)
    \end{array}\right) =
    \left(\begin{array}{c}
        x^{[i]}(z) - x^{[i]}(0) \\
        y^{[i]}(z) - y^{[i]}(0)
    \end{array}\right),
    }
\end{equation}%
where $i$ is an index arbitrarily assigned to identify each node of the FE mesh, $z$ is the position along the line, and $\big(x^{[i]}(z),\, y^{[i]}(z)\big)^\intercal$ and $\big(x^{[i]}(0),\, y^{[i]}(0)\big)^\intercal$ are the cross-sectional positions of the $i$-th node at position $z$ and at the beginning of the line, respectively. These deformations are of course affected by the uncertain (uncontrollable) and deterministic (controllable) process parameters defined in the previous section. To simplify the notation, the explicit dependence on the vectors $u$ and $c$ has been omitted from the above expression.

The TWINKLE library uses tensor factorization to obtain a description of each output as a combination of different terms which contain the product of nonlinear functions depending on each input parameter. {In particular, the x-axis deformation of the $i$-th node at position $z$ is approximated by the ROM as}
\begin{equation} \label{eq:twinkle}
{\Delta x^{[i]}(z)=\sum_{m=1}^{M} \alpha_{m}^{[i]}\, f^{[i]}_m(z) \prod_{n=1}^{3}g^{[i]}_{m,n}( {u}_{n}) \prod_{p=1}^{6}h^{[i]}_{m,p}( {c}_{p})},
\end{equation}
where $M$ is the number of terms {(approximation order), $u_n \in \mathbb{R}$ is the $n$-th element of $u$, $c_p \in \mathbb{R}$ is the $p$-th element of $c$, $\alpha_m^{[i]}$, $m=1, \dots , N$ are weighing coefficients, and $f_m$, $g_{m,n}$, and $h_{m,p}$ are one-dimensional nonlinear functions. The deformation in the y-axis, $\Delta y^{[i]}(z)$, is computed in an equivalent way.}

\subsection{Uncertainty modeling} 

The ROM previously described is used for estimating the response of the system during the optimization process. Each call to the model must be understood as a simulation that captures the properties of the material section along the line in stationary conditions, that is to say, how it behaves given a certain configuration of the process parameters in Table~\ref{ref_values}. The control is focused on the deterministic parameters, which are considered as controllable inputs. Since the production is also influenced by uncertain parameters, the same configuration for the controllable parameters might lead to different quality outputs. This effect has been modeled by randomly sampling values for the uncertain parameters each time the model is invoked. This aims to represent the fact that each time a control configuration is evaluated, it will be applied to a system with a particular realization for the uncertain parameters. 

\begin{table} [t]
    \begin{center}
    \begin{tabular}{ | p{2cm} | p{2.2cm} | p{3cm} | }
         \hline
         \textbf{Input}  & \textbf{Distribution} &  \textbf{Distribution \mbox{parameters}} \\ 
         \hline
         \mbox{Extrusion} \mbox{speed}  & \mbox{Normal} & \mbox{$\mu = 20$} \newline \mbox{$\sigma = 0.5$}  \\
         \hline
         \mbox{Microwave} \mbox{oven} \mbox{ratio}  & Normal & $\mu = 0.55$ \newline $\sigma = 0.08$ \\  
         \hline
         \mbox{Foaming} \mbox{coefficient}  & Log normal & $\mu = \log(0.08)$ \newline $\sigma = 0.262$  \\
         \hline
    \end{tabular}
    \caption{Probability distribution parameters for uncertain inputs}
    \label{tab:PDFs}
    \end{center}
\end{table}

\begin{figure}[t]
    \centering
	\includegraphics[width=85mm]{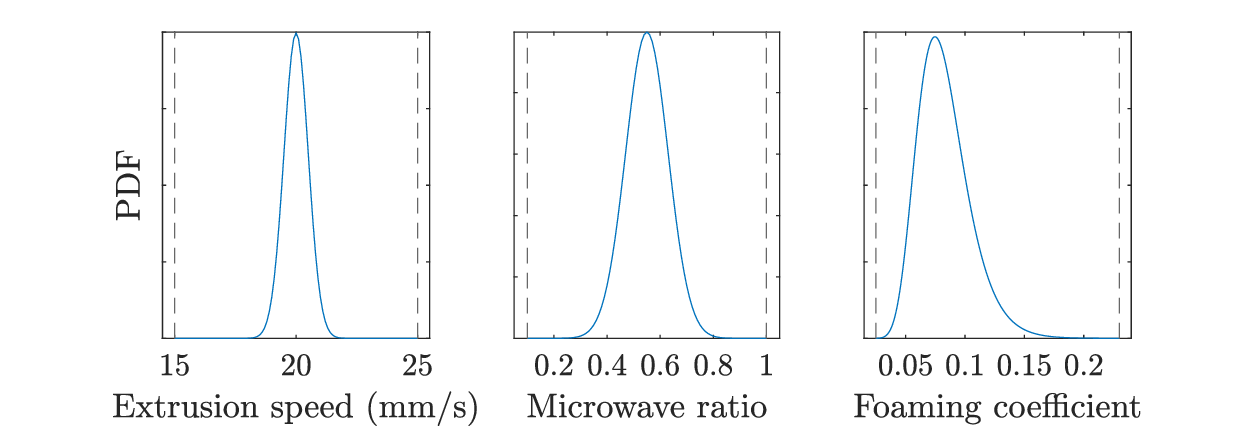}
	\caption{Probability density functions for the uncertain inputs}
	\label{pdf}
\end{figure}

The uncertain parameters correspond to external factors, such as previous processes (e.g. extrusion speed), the use of a specific material batch (e.g. foaming coefficient) or uncertainties in the system devices (e.g. microwave oven). Although they are unknown, they remain constant in each production, that is to say, in each simulation of our model. The distributions for the uncertain parameters are similar to those in~\cite{Monge2019ReducedOM}. The particular choice of probability distributions is detailed in Table \ref{tab:PDFs}. The sampling has been truncated between the percentiles that yield to the model domain extremes as it is depicted in Figure \ref{pdf}.

\section{Optimization methodology}  
\label{sec:optim}

\subsection{Objectives and formulation}

\begin{figure}[t]
    \centering
	\includegraphics[width=85mm]{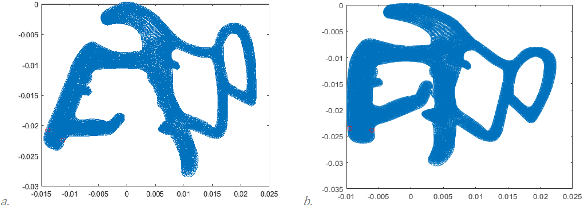}
	\caption{Variation in the geometry of the part at the beginning (a) and at the end (b) of the line}
	\label{section_change}
\end{figure}

As a consequence of the processes experienced along the line, the section shape changes as it can be seen in Figure \ref{section_change}. In order to ensure the functionality of the seal in further assembly stages, there exists a quality requirement to keep a certain dimension at the end of the line. The location of the gauged points under control are marked in red in Figure \ref{section_change}. The objective of this work is to find a configuration for the controllable parameters ($c$) that yields to a reasonable value in the controlled distance {between these two points}, despite of the uncertain factors ($u$).

The ROM predicts the geometrical evolution of the two control points of the part along the line. In particular, the model computes the deviation from the starting position for such nodes. {The in-between distance for a given position $z$ at the line, $d(z)$, can be thus computed as follows,}
\begin{equation}
    {d(z) = \lVert (d_x(z), d_y(z)) \rVert _ {2}}\, ,
\end{equation}%
\vspace{-\baselineskip}
{\begin{align}     
     d_x(z) &= x^{[i_2]}(z) - x^{[i_1]}(z)\nonumber\\
     &= x^{[i_2]}(0) + \Delta x^{[i_2]}(z) - x^{[i_1]}(0) - \Delta x^{[i_1]}(z),\\[1em]
     d_y(z) &= y^{[i_2]}(z) - y^{[i_1]}(z)\nonumber\\
     &= y^{[i_2]}(0) + \Delta y^{[i_2]}(z) - y^{[i_1]}(0) - \Delta y^{[i_1]}(z),
\end{align}}%
{where $i_1$ and $i_2$ are the identification indices of the two gauged points.}

The objective of getting a desired value for that distance at the end of the line {(position $z= z_\mathrm{f}=105.85$~m)} has been tackled via an optimization problem, in which the controllable inputs, $c$, are tuned to minimize a cost function that is larger the further the distance is from the setpoint, \textit{r}. {Mathematically, the problem is stated as follows:}
\begin{equation} \label{eq:optimization}
\min_{c} \quad J(c, u)
\end{equation}
\begin{equation}
{\textrm{s.t.} \quad c  \in \mathrm{dom}(model)}
\end{equation}
where $\mathrm{dom}(model)$ is the domain defined by the parameter ranges in table \ref{ref_values}, and the cost function is given by: 
{\begin{align} 
 J(c, u) =\, & \lVert d (z_\mathrm{f}) - r \rVert _{2}^2 + 0.5\cdot\lVert d(z_\mathrm{f}-5) - r \rVert _{2}^2 \, + \nonumber\\
&   + 0.25 \cdot \lVert d(z_\mathrm{f}-10) - r \rVert _{2}^2 +\gamma \lVert c \rVert _{2} \label{eq:cost}
\end{align}}%
This function considers the geometrical deviation at the end of the line {($z=z_\mathrm{f}$) and also at previous positions ($z=z_\mathrm{f}-5=100.85$~m and $z=z_\mathrm{f}-10=95.85$~m)} in order to improve the stability of the distance under control. In addition to that, a term penalizing the control parameters ($\gamma \lVert c \rVert _{2}$) is included for fostering the efficiency of the process as the controllable parameters are directly related to energy consumption (oven power, blowing pressures, etc.). 

The function \eqref{eq:cost} under optimization is stochastic due to the uncertain parameters ($u$) that take values at random each time it is evaluated. Given that, we seek values for the controllable parameters ($c$) that yield to a \textit{good} probability distribution of the minimized function. In particular we look for a solution that produces a distribution with the {99\%} percentile as small as possible, thus capturing an operating point that behaves good (i.e. controlled distance close to the desired value with good actuation efficiency) in most of the cases.

\subsection{Stochastic optimization algorithm} \label{subsec:algo}

The optimization problem formulated above presents two main characteristics: it has a stochastic nature and evaluations of the cost function are expensive. There exist algorithms that can efficiently deal with such formulations taking into account the nature of the function under optimization. Our choice has been the so-called Bayesian optimization. This class of algorithms are widely used and there are powerful implementations available. In particular, for this work, the optimization has been solved using the implementation of MATLAB. 

Bayesian optimization works by building a probabilistic model---a Gaussian process---that estimates the unknown function to optimize. It predicts the possible values the function could take and how uncertain are such predictions, i.e., it builds a mathematical object that estimates a probability distribution for each point in the search space (visited and not visited). {The probabilistic model starts with prior beliefs about the function's behavior based on a few initial observations, and it is refined as new points are evaluated}~\cite{Shahriari_2016}. The acquisition function for exploring the parameter domain in the present use case is "Expected-Improvement". This strategy pushes the search towards regions in which the expected reduction in the cost is the greatest. In order to limit the processing time, the stopping criterion has been set at 100 evaluations. In our case, the optimal solution is chosen as a point, which could have been visited or not, that offers a distribution with the minimum 99th percentile.

Despite of the clear advantages of this method, it still makes some assumptions that are not fully satisfied in our problem. The MATLAB implementation assumes that the probability density functions are Gaussian, which in reality is not true. The cost function is non-symmetric, and also constrained by a lower bound at 0. In addition to that, the strategy to build the probabilistic model uses kernels that are also based on Gaussian properties. Bearing in mind such limitations, we decided to repeat the algorithm one hundred times, that is to say, initializing the probabilistic model with different initial evaluations. This strategy aims to be robust by avoiding the choice of solutions that could correspond to local optima. Among the hundred different solutions, we have chosen as the best optimizer the one that shows the minimum 99th percentile in the cost. The percentile evaluation has been done on sample based probability functions computed with 10,000 evaluations, in which the controllable parameters are provided by the Bayesian optimization process and the non-controllable ones take values at random.

\section{Results}
\label{sec:results}

\subsection{Stochastic optimization repeatability}

\begin{figure*}[ht]
    \centering
    \includegraphics[width=\linewidth]{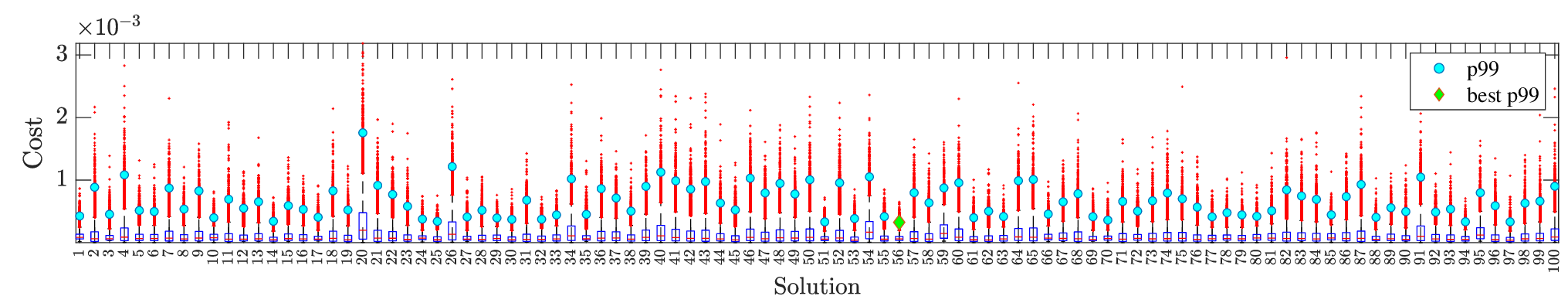}
	\caption{Cost distribution box plots of the Bayesian Optimization solutions. Percentile 99 highlighted in cyan, best solution marked in green.}
	\label{boxplot}
\end{figure*}

This section presents the results of the stochastic optimization strategy described in Section \ref{subsec:algo}. For each of the 100 solutions for the controllable inputs previously found, the cost function has been evaluated 10,000 times in order to show, by sampling, the approximate probability distribution of such cost function given a particular solution for the controllable inputs and letting the uncertain inputs take values at random following the distributions presented earlier. {Figure \ref{boxplot} displays the main results of this procedure through boxplots, with each solution's 99th percentile indicated by a cyan dot.}

Although in general the 99th percentile is around $J=10^{-3}$, some disparities can be seen in the distributions. There are some processes, such as number 20, that found a solution with a really high percentile, indicating that the Bayesian search was no successful at all. On the contrary, there are other times where the 99th percentile is even lower than $0.5 \cdot 10^{-3}$. 
Among those solutions that give rise to a favorable distribution, number 56 is the one with the lowest 99th percentile. {That percentile in that particular solution is highlighted with a green dot in Figure \ref{boxplot}}. As stated in the methodology chapter, such solution is chosen as the best optimizer for our problem.  

\subsection{Comparison with a deterministic strategy}

Here we present a comparison between the solution obtained with the methodology presented above and one obtained using the Simplex optimization algorithm over the same problem formulation. Given the deterministic nature of the algorithm, a lower performance is expected given the inherent variability of the process and the incapability of the optimization algorithm to take it into account. In consequence it is prone to get trapped in local minima points. As explained in~\cite{1991_Barton} the Nelder-Mead optimization algorithm, which is broadly used, tends to prematurely terminate in presence of large enough stochastic noise.

\begin{figure}[t]
    \centering
	\includegraphics[width=85mm]{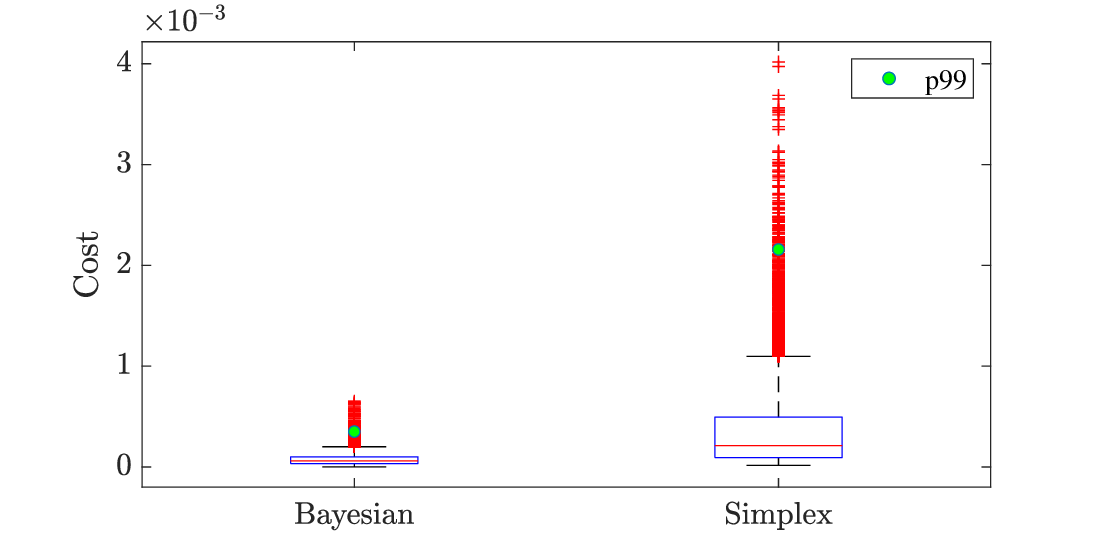}
	\caption{Box plots comparing the cost distribution in the optimal solution for Bayesian and Simplex algorithms}
	\label{benchmarking}
\vspace{\floatsep}
    \centering
	\includegraphics[width=85mm]{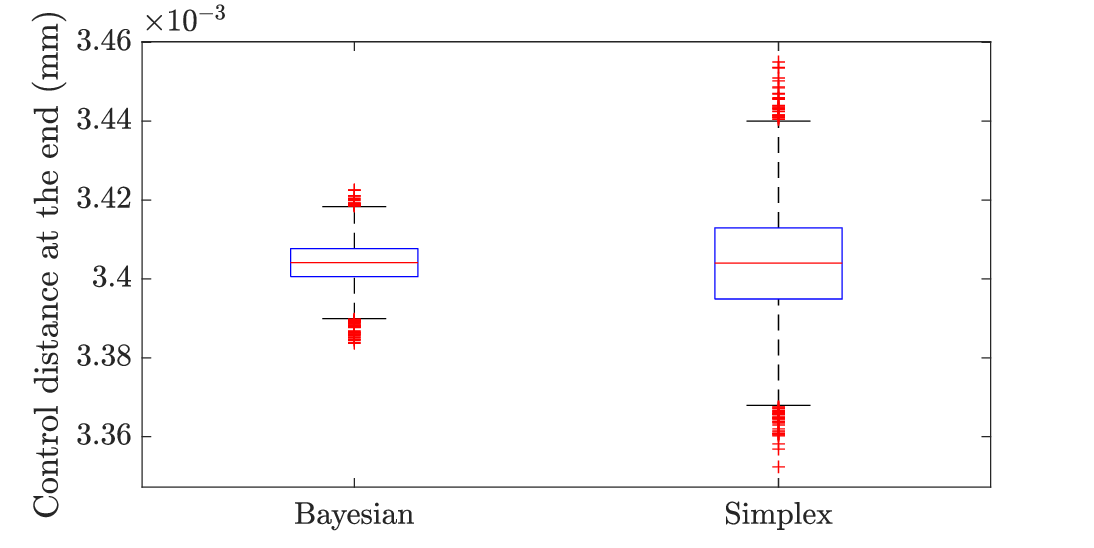}
	\caption{{Box plots comparing the control distance distribution at the end of the line for the optimal solution for Bayesian and Simplex algorithms}}
	\label{final_slices}
\end{figure}

To compare the quality of the optimal solutions provided by these two methods, the distributions for the cost function and the controlled distance have been evaluated using a Monte Carlo simulation. A total of 10,000 triplets for the uncertain parameters have been sampled from the distributions shown in Figure \ref{pdf}, resulting in 10,000 values for the metrics under study. Figure \ref{benchmarking} shows box plots comparing the cost distribution around the best solution from the Bayesian optimization and around the optimal solution from the Simplex. On the other hand, Figure \ref{final_slices} shows the distributions of the controlled distance at the end position ($z=z_\mathrm{f}$) of the production line when the controllable parameters are set to the optimal values obtained from the corresponding optimizations.

Although the Simplex solution is not bad, the Bayesian optimization solution outperforms it, shrinking the probability distribution to smaller values. The 99th percentiles are depicted with green dots, clearly showing a smaller value in the Bayesian optimization strategy. {Additionally, the controlled distance also depicts a tighter distribution around the target (3.4 mm) in the case of the Bayesian methodology. This is, after all, the performance metric that is optimized in this work.}

\subsection{{Further results and discussion}}

\begin{figure}[t]
    \centering
	\includegraphics[width=85mm]{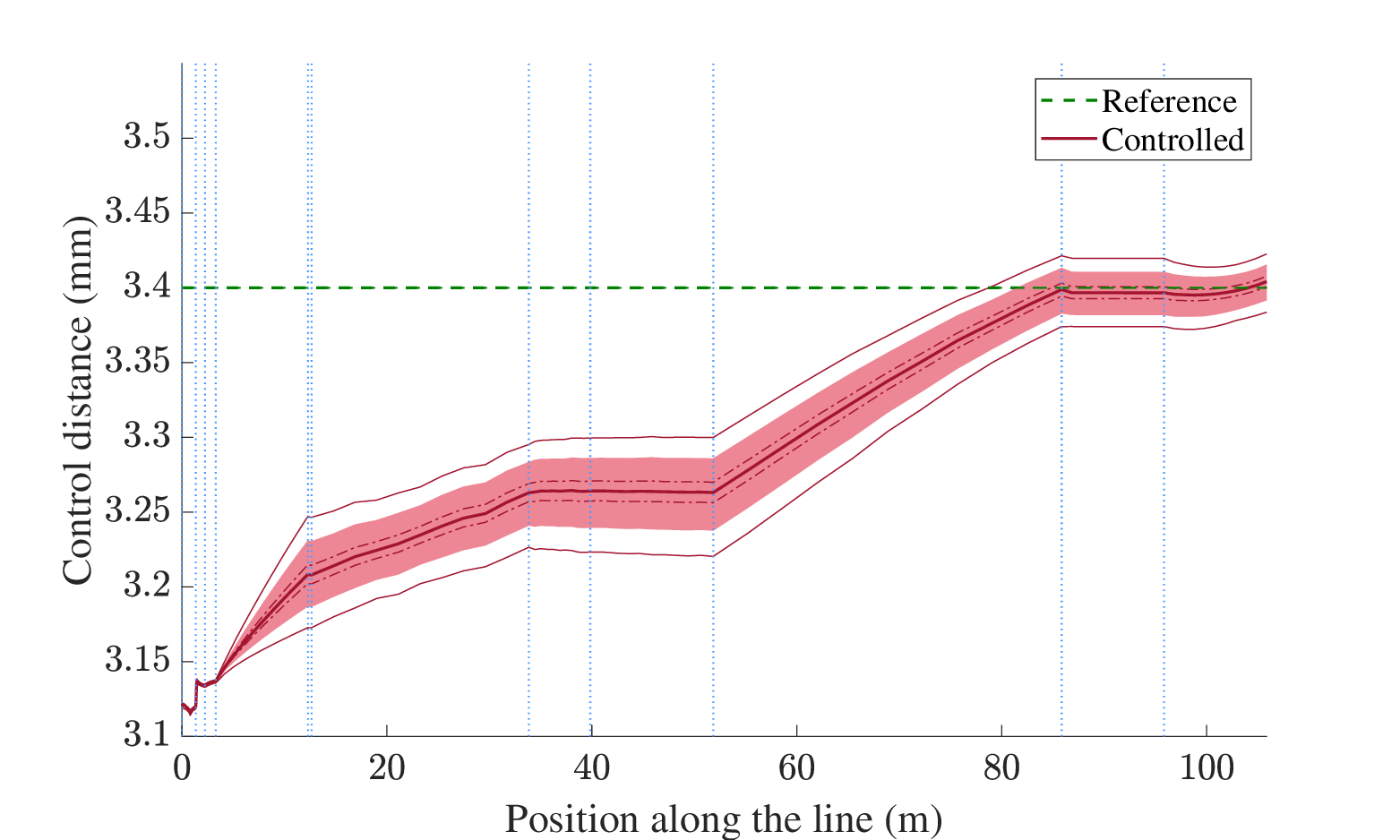}
	\caption{Controlled distance evolution along the line. Outerlines: maximum and minimum values, shaded: 1st-99th percentile, dotted: 25th-75th percentile, continuous: median}
	\label{lineTrajectories}
\end{figure}

After obtaining the best solution, we have tested its performance in additional simulations. We have computed 10,000 trajectories of the controlled distance along the line, while keeping the controllable parameters constant at the optimal value obtained with the Bayesian method. The uncertain parameters, on the other hand, have been assigned random values for each trajectory. For consistency, we have used the same 10,000 random triplets generated earlier to build the boxplots in Figures~\ref{benchmarking} and~\ref{final_slices}. The obtained data have been then sorted and represented in a percentile band fashion, as shown in Figure~\ref{lineTrajectories}. In this figure, the outer continuous lines represent the maximum and minimum values, i.e., all trajectories fall within these limits. The shaded area represents the range between the 1st and 99th percentiles, thus encompassing 98\% of the trajectories. The dotted lines and the thick continuous line indicate the 25th-75th percentile range and the median, respectively. The results in the figure clearly show that the controlled distance reaches, on average, the desired target at the end of the line despite the uncontrollable random parameters. It is worth noting that the distribution of distance values along the line is not monotonically increasing, as one might expect. Initially, there is no variability since all solutions start from the same value. The variability then increases and reaches a maximum at around $z=50$~m. From this point onwards, however, the distribution narrows until the end of the line, which is the position of interest as it corresponds to the finished product. As a result of the procedure followed, the distribution shown in the final position corresponds exactly to the left one in Figure~\ref{final_slices}.

The results shown in the previous figures can also be analyzed in terms of the expected quality of the final product. In this sense, the solution obtained by the Bayesian optimization method would more than guarantee a dimensional tolerance of $\pm0.03$~mm around the target dimension, regardless of the value of the uncertain factors. However, the process parameters obtained with the deterministic optimization method only ensure a tolerance of $\pm0.06$~mm around the target. It is important to note that this improvement does not require any change in the production line, but is simply the result of our optimization procedure, i.e., it does not imply any additional cost. The benefit of our proposal is thus clearly demonstrated.

\section{Conclusions}
\label{sec:conclusions}

In this article we have presented a strategy to optimize the operation of a rubber seals production line for the automotive industry. {As explained, it is technically and economically impossible to perform the optimization by acting directly on the plant. For this reason, the optimization has been performed using model-based techniques. In particular, it has been carried out by means of a simplified model obtained from a detailed finite element model. This reduces the computational load, allowing us to run thousands of simulations in different scenarios in a reasonable amount of time. In contrast, these simulations would have required unmanageable computational times if the high-order model had been used directly.}

The production line has a partially random behavior that has been also captured in the simplified model. This has allowed us to address the optimization in a robust manner using a probabilistic approach. Specifically, our proposal is based on a Bayesian optimization algorithm, and has been designed with the objective of finding the parameter configuration that gives rise to the best performance in 99\% of the cases. The presented results show that this method outperforms a deterministic algorithm, namely the simplex algorithm, which highlights the benefits of our approach.

In summary, the main advantages of our methodology are the low computational cost, associated with the use of a reduced order model, and the stochastic nature of the optimization, which allows us to obtain a robust strategy valid for virtually all the possible values of the uncontrolled variables. The method is particularly convenient if a detailed high-order model of the system is already available. Otherwise, the need to build such a model could be considered a drawback. Future work would go in two directions. On the one hand, to evaluate the possibility of building the simplified model directly from real plant data, avoiding the need to build the high-order model. On the other hand, to apply the results in the real plant and validate the proposal with real data.

\section*{Acknowledgement}

This work has been partially funded by the Government of Aragon, via grants to promote the research activity of research groups (T45\_23R and T73\_23R), and partially by the European Regional Development Fund (ERDF).
The authors would like to thank the firm Standard Profil for the possibility to use their line as reference for the presented development, and also Ismael Viejo for his support in using the TWINKLE process model.

\vspace{1em}

The authors declare that they have no known competing financial interests or personal relationships that could have appeared to influence the work reported in this paper.

\end{document}